\documentclass[12pt,epsf]{article}
\usepackage{amsmath,amssymb,graphicx,epsfig,latexsym}
\numberwithin{equation}{section}
\newcommand{\be}{\begin{equation}}
\newcommand{\ee}{\end{equation}}

\newcommand{\beqa}{\begin{eqnarray}}
\newcommand{\eeqa}{\end{eqnarray}}
\newcommand{\nn}{\nonumber}

\def\boxit#1{\vbox{\hrule\hbox{\vrule\kern8pt
\vbox{\hbox{\kern8pt}\hbox{\vbox{#1}}\hbox{\kern8pt}}
\kern8pt\vrule}\hrule}}
\def\mathboxit#1{\vbox{\hrule\hbox{\vrule\kern8pt\vbox{\kern8pt
\hbox{$\displaystyle #1$}\kern8pt}\kern8pt\vrule}\hrule}}

\def\IB{\relax\hbox{$\inbar\kern-.3em{\rm B}$}}
\def\IC{\relax\hbox{$\inbar\kern-.3em{\rm C}$}}
\def\ID{\relax\hbox{$\inbar\kern-.3em{\rm D}$}}
\def\IE{\relax\hbox{$\inbar\kern-.3em{\rm E}$}}
\def\IF{\relax\hbox{$\inbar\kern-.3em{\rm F}$}}
\def\IG{\relax\hbox{$\inbar\kern-.3em{\rm G}$}}
\def\IGa{\relax\hbox{${\rm I}\kern-.18em\Gamma$}}
\def\IH{\relax{\rm I\kern-.18em H}}
\def\IK{\relax{\rm I\kern-.18em K}}
\def\IL{\relax{\rm I\kern-.18em L}}
\def\IP{\relax{\rm I\kern-.18em P}}
\def\IR{\relax{\rm I\kern-.18em R}}
\def\IZ{\relax\ifmmode\mathchoice
{\hbox{\cmss Z\kern-.4em Z}}{\hbox{\cmss Z\kern-.4em Z}}
{\lower.9pt\hbox{\cmsss Z\kern-.4em Z}} {\lower1.2pt\hbox{\cmsss
Z\kern-.4em Z}}\else{\cmss Z\kern-.4em Z}\fi}

\def\II{\relax{\rm I\kern-.18em I}}

\def\CD {{\cal D}}

\def\CP {{\cal P}}

\begin{document}

\setlength{\baselineskip}{7mm}
\begin{titlepage}

\begin{flushright}

{\tt NRCPS-HE-47-2013} \\
{\tt CERN-PH-TH/2013-055}\\

\end{flushright}

\vspace{1cm}
\begin{center}

{\Large \it Extension of Chern-Simons Forms
\\
and
\\
New Gauge Anomalies \\
\vspace{0,3cm}

} 

\vspace{1cm}
{\sl Ignatios Antoniadis$^{~\star,}$\footnote{${}$ On leave of absence
from CPHT {\'E}cole Polytechnique, F-91128, Palaiseau Cedex,France.}
 and  George Savvidy$^{\star,+}$

\bigskip
\centerline{${}^\star $ \sl Department of Physics, CERN Theory Division CH-1211 Geneva 23, Switzerland}
\bigskip
\centerline{${}^+$ \sl Demokritos National Research Center, Ag. Paraskevi,  Athens, Greece}
\bigskip
}
\end{center}
\vspace{25pt}

\centerline{{\bf Abstract}}

\vspace{10pt}

\noindent

We  present a general analysis of  gauge invariant,   exact and
metric independent forms which can be constructed using higher rank field-strength tensors.
The integrals of these forms over the corresponding space-time coordinates
provides new topological Lagrangians. With these Lagrangians one can define
gauge field theories which generalize  the Chern-Simons quantum field theory.
We also present explicit expressions for the potential gauge anomalies associated
with the tensor gauge fields and classify all possible anomalies
that can appear in lower dimensions. Of special interest are those which can be
constructed in four, six, eight and ten dimensions.

\begin{center}
\end{center}

\vspace{150 pt}


\end{titlepage}

\newpage

\pagestyle{plain}

\section{\it Introduction}

The chiral anomalies appear
in gauge theories interacting with Weyl fermions.
The $U_A(1)$  gauge anomaly is given by the  Pontryagin-Chern-Simons
 $2n$-form
\cite{Zumino:1983ew,Stora:1983ct,Faddeev:1984jp,Faddeev:1985iz,LBL-16443,Manes:1985df,Treiman:1986ep,Faddeev:1987hg,AlvarezGaume:1985ex}
\be\label{abeliananom}
d*J^A ~\propto ~\CP_{2n}=Tr(G^n)= d ~\omega_{2n-1},
\ee
where $\omega_{2n-1}$ is  the  Chern-Simons form in $2n-1$ dimensions
\cite{Zumino:1983ew,Treiman:1986ep}:
\be\label{integralformforAbelian}
\omega_{2n-1}(A)= n \int^1_0 d t ~ Tr(AG^{n-1}_t)\, .
\ee
$G=dA +A^2$ is the 2-form  Yang-Mills (YM) field-strength tensor
of the 1-form vector field\footnote{$L^a$ are the generators of the Lie algebra.}  $A = -ig A^{a}_{\mu} L^a dx^{\mu}$ and
$G_{t}= t G +(t^2-t)A^{2}$. A celebrated result for the non-Abelian anomaly
\cite{Adler:1969gk,Bell:1969ts,Bardeen:1969md,Wess:1971yu,
Frampton:1983nr} can be obtained by
gauge variation of the $\omega_{2n-1}$
\cite{Zumino:1983ew,Stora:1983ct,Faddeev:1984jp,Faddeev:1985iz,LBL-16443,Treiman:1986ep,Faddeev:1987hg,AlvarezGaume:1985ex}:
\be\label{definition}
 \delta    \omega_{2n-1} = d \omega^1_{2n-2}~,
\ee
where the $(2n-2)$-form has the following integral representation of Zumino \cite{Zumino:1983ew}:
\beqa\label{celebratedanomaly}
\omega^{1}_{2n-2}(\xi, A)=n(n-1) \int^{1}_{0} d t (1-t)~
Str \left(  \xi  d (A ~G^{n-2}_{t}  ) \right)\, .
\eeqa
Here, $\xi = \xi^a L_a$ is a scalar gauge parameter and $Str$ denotes a symmetrized trace. The
covariant divergence of the non-Abelian
left and right handed currents is given by this $(2n-2)$-form:
\be
D*J^{L,R}_{\xi} ~\propto    ~\omega^{1}_{2n-2}(\xi,A).
\ee
Thus the non-Abelian anomaly  in $2n-2$-dimensional space-time may be obtained  from
the Abelian anomaly (\ref{abeliananom}) in $2n$ dimensions by a series of reduction (transgression) steps
(\ref{abeliananom}) and (\ref{definition})
by a differential geometric method without having to evaluate the
Feynman diagrams \cite{Zumino:1983ew,Stora:1983ct,Faddeev:1984jp,Faddeev:1985iz,LBL-16443,
Manes:1985df,Treiman:1986ep,Faddeev:1987hg,AlvarezGaume:1985ex,Adler:1969gk,Bell:1969ts,Bardeen:1969md,Wess:1971yu,
Frampton:1983nr,Mickelsson:1983xi,Faddeev:1986pc,DKFaddeev,Grimm:1974pr,Bardeen:1984pm,
AlvarezGaume:1984dr,AlvarezGaume:1985yb}.

In recent articles \cite{arXiv:1001.2808,Antoniadis:2012ep,Georgiou:2012fr} the authors
have found similar invariants in
non-Abelian tensor gauge field theory \cite{Savvidy:2005fi,Savvidy:2005zm,Savvidy:2005ki}.
The first series of exact $(2n+3)$-forms are defined as follows:
\be\label{oldanomaly}
\Gamma_{2n+3} = Tr(G^n G_3) = d \sigma_{2n+2},
\ee
where $G_3=dA_2 +[A,A_2]$ is the 3-form  field-strength tensor
for the rank-2 gauge field
$A_2= -ig A^{a}_{\mu\lambda} L_a dx^{\mu} \wedge dx^{\lambda}$.
The second series of exact $6n$-forms is defined as \cite{Antoniadis:2012ep}:
\be
\Delta_{6n} = Tr(G_3)^{2n}= d \pi_{6n-1},
\ee
where  the $(6n-1)$-forms are defined in $\CD=6n-1 $ dimensions.
The third series of invariant forms are defined in  $\CD=2n+4 $ dimensions and are given by the expression
\be\label{generalized}
\Phi_{2n+4} =Tr(G^n  G_4) = d \psi_{2n+3}~,
\ee
where the corresponding secondary $(2n+3)$-form $\psi_{2n+3}$ is defined in $\CD=2n+3 $ dimensions
and $G_4 =d A_3 + \{ A, A_3\}$.
The forth series of forms is defined in $\CD=2n+2$ dimensions~\cite{Georgiou:2012fr}
\be\label{linearcombination}
\Omega_{2n+2} = Tr(G G_{2n} + ....)= d \chi_{2n+1}
\ee
where the forms  $\chi_{2n+1}$ are defined
in  $\CD=2n+1 $ dimensions.
All new forms $\Gamma_{2n+3} $, $\Delta_{6n} $, $\Phi_{2n+4} $ and
$\Omega_{2n+2} $ are analogous to the   Pontryagin-Chern-Simons densities
$\CP_{2n}$  in YM gauge theory (\ref{abeliananom}): they are {\it gauge invariant, exact and metric independent}.
Our aim is to find out all potential gauge anomalies which are
generated by the new invariant forms in even dimensions performing transgression
analogous to the (\ref{abeliananom}) and (\ref{definition})
\be
\CP_{2n}~ \Rightarrow ~\omega_{2n-1} ~\Rightarrow ~\omega^{1}_{2n-2}.
\ee
In particular we are interested to enumerate
and classify all potential  anomalies and their structure in even dimensions from four to ten
space-time dimensions.
For this purpose one should enumerate all gauge invariant, exact and metric independent
forms in the corresponding dimensions. The integrals of these forms over the corresponding space-time coordinates
provides us with new topological Lagrangians and with a generalization of the
Chern-Simons quantum field theory \cite{ Witten:1988hf,Schwarz:1978cm,Schwarz:1978cn,Schonfeld:1980kb,Deser:1982vy,
Deser:1981wh,Witten:1992fb, Beasley:2005vf, Witten:1989ip,
Axelrod:1989xt, Dijkgraaf:1989pz}.

In the next Section~2, we shall present a general analysis of  such forms
which can be constructed using higher rank field-strength tensors, and we will define
gauge field theories that generalize  the Chern-Simons quantum field theory. In
Section~3, we will present explicit expressions for the potential gauge anomalies associated
with the tensor gauge fields. They are given by the sum of the two formulas (\ref{resultmain})
and (\ref{tensoranomaly}) representing our result for the gauge anomalies.
From this calculation it turns out that
at least two powers of the YM field-strength tensor $G$ should be present in the
form $\Phi_{2n}$ to generate anomalies with respect to the standard
gauge transformations.
In Section~4, we shall classify all possible forms
which can be constructed in lower dimensions. Of special interest are those that can be
constructed in four, six, eight and ten dimensions. Finally, in the Appendix, we present useful formulas
for the gauge transformations of fields, the corresponding Biachi identities and for a one parameter
deformation of fields generalizing Zumino's construction.

\section{\it Gauge Invariant, Exact and Metric Independent Forms }

In order to completely enumerate and classify  all possible gauge invariant, exact and metric independent forms
in extended YM theory we shall start from the Pontryagin-Chern-Simons density  $\CP_{2n}$  in
YM theory and then proceed with a number of steps by decreasing the power of YM field-strength tensor $G$ and
increasing the power of the higher rank field-strength tensors $G_{2n}$, but keeping the rank of the forms  fixed.
Thus, we have to study the following sequence of $2n$-forms
\beqa
&&Tr(G^{ n}),\nn\\
&&Tr(G^{ n-2} G_4 ),\nn\\
&&Tr(G^{ n-3} G_6 ),\nn\\
&&Tr(G^{ n-4} G_8 ),~~~ Tr(G^{ n-4} G^2_4 ),\nn\\
&&Tr(G^{ n-5} G_{10} ),~~~ Tr(G^{ n-5} G_6 G_4 ),\nn\\
&&Tr(G^{n-6} G_{12} ),~~~ Tr(G^{ n-6} G_8 G_4 ),~~~ Tr(G^{ n-6} G^3_4 ),\nn\\
&&Tr(G^{n-7} G_{14} ),~~~ Tr(G^{ n-7} G_{10} G_4 ),~~~ Tr(G^{ n-7} G_8 G_6 ),\nn\\
&&Tr(G^{n-8} G_{16} ),~~~ Tr(G^{ n-8} G_{12} G_4 ),~~~ Tr(G^{ n-8} G_{10} G_6 ),~~~ Tr(G^{ n-8} G_{8} G_8 ),\nn\\
&&~~~~~~~~~~~~~~~~~~~~~~~ Tr(G^{ n-8} G_8 G^2_4 ),~~~ Tr(G^{ n-8} G^2_6 G_4),~~~ Tr(G^{ n-8} G^4_{4}  ),\nn\\
&&..............................
\eeqa
By construction they
are metric independent, and thus invariant under general coordinate transformations because
the indices of the field-strength tensors are contracted  with the totally antisymmetric Levi-Civita
tensor $\epsilon^{\mu_1,...,\mu_{2n}}$.
As a next step, one should check that these forms are gauge invariant and exact. Not all of them
share  these properties and as we shall see in some cases one should consider linear
combinations of these forms as the $\Omega_{2n } $ in (\ref{linearcombination}).
Those of the  forms which will be found to be gauge invariant and exact,
by transgression will generate  potential non-Abelian anomalies in $2n-2$ dimensions.

The forms $\Gamma_{2n+3} $, $\Delta_{6n} $, $\Phi_{2n+4} $ and
$\Omega_{2n+2} $ which we already presented in the introduction are part of the
above list and are gauge invariant, exact and metric independent. For instance, the series
of forms linear in $G_6$ can be easily constructed because they contain only the lower
field-strength tensor $G_4$. The first forms are:
\beqa
\Xi_{8}&=& Tr(G  G_6 + G_4 G_4)  = d \phi_7 ,\nn\\
\Xi_{10} &=& Tr(G^2 G_6 +2 G G^{2}_{4})  =d \phi_{9},\\
\Xi_{12} &=& Tr(G^3 G_6 +2 G^2 G^2_{4}  + G G_4 G G_4)  =d \phi_{11}\nn\\
.........& &..............................
\eeqa
which can be written using the symmetrized trace, as
\be
\Xi_{2n+6}=Str(G^n G_6 +n G^{n-1}  G^2_{4}  )  =d \phi_{2n+5}.
\ee
The lower dimensional forms linear in $G_8$ are
\beqa
\Upsilon_{10}&=& Tr(G  G_8 + 3 G_4 G_6)=d \varrho_9 ,\nn\\
\Upsilon_{12} &=& Tr(G^2 G_8 +3 G  G_{4} G_6 + 3 G  G_6 G_{4} +2  G^3_4  )  =d \varrho_{11},\nn\\
...........& &.................................
\eeqa
and so on.

The general analysis of possible forms will be presented in the following sections,
but already at this stage one can see that there is a reach class of invariant densities
which are relevant for the description of possible gauge anomalies.
At the same time, integrals of these forms over the corresponding space-time coordinates
provides us with new topological Lagrangians. In particular, one can define
topological
field theories, generalizing the Chern-Simons quantum field theory
\cite{ Witten:1988hf,Schwarz:1978cm,Schwarz:1978cn,Schonfeld:1980kb,Deser:1982vy,
Deser:1981wh,Witten:1992fb, Beasley:2005vf, Witten:1989ip,
Axelrod:1989xt, Dijkgraaf:1989pz}, in which the
correlation functions have support on two-dimensional surfaces $M^i_2$ and knots $C^j$
\beqa
Z(M,M^i_2,C^j,R) = \int \CD A \CD A_2  e^{ i k \int_{M } \sigma_{2n+2}(A,A_2) }
\prod_{i,j} Tr_{R_i} e^{  i   \oint_{M^i_2} A_2   }  Tr_{R_j} e^{  i   \oint_{C^j } A    },
\eeqa
where $\sigma_{2n+2}$ is defined in (\ref{oldanomaly}) and k is a parameter,
or on three-dimensional manifolds
\beqa
Z(M,M^i_3,C^j,R) = \int \CD A \CD A_3 e^{i k \int_{M } \psi_{2n+3}(A,A_3)} \prod_{i,j} Tr_{R_i}
e^{  i   \oint_{M^i_3} A_3   }  Tr_{R_j} e^{  i   \oint_{C^j } A    }
\eeqa
as well as on higher dimensional ones, $ \psi_{2n+3}$ is defined in (\ref{generalized}). In particular, for the partition function $Z(M)$ in four dimensions
\cite{arXiv:1001.2808,Antoniadis:2012ep}, we get
\be\label{sigma4}
Z(M ) = \int \CD A \CD A_2 e^{i k \int_{M_4} \sigma_{4}} = \int \CD A \CD A_2 e^{i k \int_{M_4}  Tr( G A_2)}
\ee
and in the large k limit the contribution to the path integral is dominated
from the points of stationary phase which are, in the given case, the flat connections
\be
G=dA +A^2=0,~~~~G_3=dA_2 +[A,A_2]=0.
\ee
The solutions of the first equation are well known $A^{flat}=g^{-1}dg$, while the
solutions of the second one have been found in \cite{Antoniadis:2012ep}
$$
A^{flat}_2=g^{-1}dg_1 -g^{-1}g_1g^{-1} dg.
$$
With these solutions in hands one can calculate the Gaussian integrals in (\ref{sigma4})
and express  the partition function $Z$ in terms of
determinants of certain operators.
The details will be given elsewhere.

\section{\it Forms $\Phi_{2n}$ Linear in $G_4$ and $\Omega_{2n}$ Linear in $G$}

We shall start our analysis with the second form in our list $\Phi_{2n} = Tr(G^{ n-2} G_4 )$,
which is the next to the standard
Pontryagin-Chern-Simons density  $Tr(G^{ n}) $ and is its natural generalization. The form  $\Phi_{2n}$ was
already considered in \cite{Georgiou:2012fr} and was shown to be gauge invariant and exact
\beqa
d \Phi_{2n}  =0.
\eeqa
According to Poincar\'e's lemma, this equation implies that $\Phi_{2n}$ can be
locally written as an exterior differential of a certain $(2n -1)$-form
\be\label{result}
Tr(G^{n-2}  G_4) = d \psi_{2n-1}~.
\ee
Generalizing  Zumino's construction \cite{Zumino:1983ew},
a one-parameter family of potentials and strengths
should be introduced:
\be
A_{t}= t A,~~~G_{t}= t G +(t^2-t)A^{2},~~~
A_{3t}= t A_3,~~~G_{4t}= t G_4 +(t^2-t)\{A,A_{3}\},
\ee
where the parameter t runs in the interval ($0\leq t \leq 1$),
so that the corresponding secondary $(2n-1)$-form is given by~\cite{Georgiou:2012fr}
\be\label{secondaryformnew}
\psi_{2n-1}(A,A_3) =  \int^1_0 d t ~Tr(A  G^{n-3}_{t} G_{4t} +...
+G^{n-3}_{t}  A G_{4t} + G^{n-2}_{t} ~ A_3).
\ee
The lower dimensional forms are:
\beqa\label{newsecondary}
\psi_{5}&=& Tr( G A_3),\nn\\
\psi_{7} &=&{1\over 3} Tr( A  G  G_4 + A  G_4 G   + A_3 G^{2}
- {1\over 2} A^3  G_4\nn\\
&-&{1\over 2} (A^2  A_3 +A A_3 A  + A_3 A^2 )G  + {1\over 2} A^4  A_3 ),\nn\\
&&......................................................
\eeqa
In order to calculate the variation of the secondary characteristics  $\psi_{2n-1}$
we need the gauge transformations of the various fields involved
in the expressions (\ref{secondaryformnew}) for $\psi_{2n-1}$,
which read
\beqa\label{gaugevariation}
&\delta A = d\xi + [A,\xi],~~~~~\delta A_3 = d\zeta_2 + [ A, \zeta_2 ] + [A_3,\xi] ,~~~\nn\\
&\delta_{\xi}d A =  [d A,\xi]- \{ A , d\xi \}  ,~~~~~\delta d A_3 =  [d A,\zeta_2]-
\{ A, d \zeta_2 \} + [d A_3, \xi] - \{ A_3, d \xi \},~~~\nn\\
&\delta_{\xi} G_t = [G_t,\xi] +(t^2-t) \{ A , d \xi \},~\nn\\
&\delta  G_{4t} = [G_{4t},\xi]+[G_t,\zeta_2] +(t^2-t)( \{A, d\zeta_2 \} +\{ A_3, d \xi \}),
\eeqa
where $\zeta_2(x) = \zeta^a_{\sigma_1 \sigma_2}(x) L_a dx^{\sigma_1} \wedge dx^{\sigma_2}$
is the rank-2 tensor gauge parameter.
It is difficult to perform in a straightforward way the variation of the $\psi_{2n-1}$,  instead one
should try to represent it in terms of the symmetrized traces\footnote{One should keep in mind that not
all expressions can be represented in the form of symmetrized traces.}. The fact that the trace in
(\ref{secondaryformnew}) can be transformed into the symmetrized trace can be proven by performing
a cyclic permutation of all terms in the integrant and then combining them into one term symmetric under
all permutations. As a result, we get the following expression
\be
\psi_{2n-1}  =  \int^1_0 d t ~Str\left[ (n-2) A  G^{n-3}_{t} G_{4t}+ G^{n-2}_{t} ~ A_3\right].
\ee
Because we have two independent gauge parameters $\xi(x)$ and $\zeta_2(x)$, we can perform the variation
of $\psi_{2n-1}$ over these gauge transformations independently.

First we shall perform the variation over
the tensor gauge parameters $\zeta_2(x)$. The terms linear in $\zeta_2(x)$ cancel out and remain
only terms linear in its differential $d \zeta_2$
\be
\delta_{\zeta} \psi_{2n-1}  =  \int^1_0 d t ~Str\left[ (n-2)(t^2 -t) A  G^{n-3}_{t} \{A, d\zeta_2 \}+ G^{n-2}_{t} ~ d\zeta_2\right].
\ee
Opening the bracket and rearranging the terms, one obtains
\beqa
\delta_{\zeta} \psi_{2n-1}  =  \int^1_0 d t ~Str\left[ (n-2)(t  -1)(~ t \{A,A\}d\zeta_2 G^{n-3}_{t}+
A d\zeta_2 [A_t,G^{n-3}_t] ) + G^{n-2}_{t} ~ d\zeta_2\right] \nn
\eeqa
and then using the equations
\be\label{helpfull}
dG^{n-2}_{t}= -[A_t,G^{n-2}_t],~~~~{\partial G_t \over \partial t}=dA +t \{A,A\}
\ee
we get
\beqa\label{intermidea}
\delta_{\zeta} \psi_{2n-1}  =  \int^1_0 d t ~Str\left[ -(n-2)(t  -1)(~ d A d\zeta_2 G^{n-3}_{t}+
A d\zeta_2 d G^{n-3}_t] \right] + \nn\\
+(n-2)(t-1)G^{n-3}_{t}{\partial G_t \over \partial t}~ d\zeta_2  + G^{n-2}_{t} ~ d\zeta_2)=\nn\\
=(n-2) \int^1_0 d t (1-t) ~Str\left[ d (G^{n-3}_{t}  A) ~d\zeta_2 \right] = d \psi^1_{2n-2},
\eeqa
where the last two terms in the second line cancel out after partial integration. Thus we arrive to the final result
\beqa\label{tensoranomaly}
\psi^{1}_{2n-2}(\zeta_2,A)=(n-2)  \int^{1}_{0} d t  (1-t)  ~Str\left(\zeta_2 ~d ~( G^{n-3}_{t} A) \right)  .
\eeqa
These forms describe the potential gauge anomalies with respect to the tensor gauge transformations
induced by the rank-2 gauge parameter   $\zeta_2(x)$.

In order to calculate the variation of $\psi_{2n-1}$ with respect to the standard gauge
transformation parameter $\xi(x)$ we should use again formulas (\ref{gaugevariation})
\beqa
\delta_{\xi} \psi_{2n-1} & =&  \int^1_0 d t Str\left[ (n-2) d \xi G^{n-3}_{t} G_{4t}
+(n-2)(n-3)(t^2 -t) A  \{A, d\xi \} G^{n-4}_{t} G_{4t} +\nn\right.\\
&+&\left. (n-2) (t^2 -t) A  G^{n-3}_{t}  \{A_3, d\xi \}
+(n-2)(t^2 -t)  \{A, d\xi \} G^{n-3}_{t} ~ A_3\right],
\eeqa
where the terms linear in $\xi(x)$ cancel out and remain
only terms linear in its differential $d \xi(x)$. Using equations (\ref{helpfull}) and
\be\label{helpfull1}
dG_{4t}= -[A_t,G_{4t}]-[A_{3t},G_{t}],~~~~{\partial G_{4t} \over \partial t}=dA_3 +2 t \{A,A_3\}
\ee
one can transform the above variation into a form similar to (\ref{intermidea})
\beqa
\delta_{\xi} \psi_{2n-1} =  (n-2)\int^1_0 d t (1-t) Str\left[ d((n-3)A G^{n-4}_{t} G_{4t}  +
   G^{n-3}_{t}  A_3 )~ d\xi  \right]= d \psi^{1}_{2n-2} ,\nn
\eeqa
and thus we get the additional gauge anomaly generated by the tensor gauge field $A_3$
\be\label{resultmain}
\psi^{1}_{2n-2}(\xi,A, A_3)=(n-2)\int^1_0 d t (1-t) Str\left[ \xi ~ d(~ (n-3)G^{n-4}_{t} G_{4t} A +
   G^{n-3}_{t}  A_3 )\right].
\ee
The sum of the two expressions (\ref{resultmain}) and (\ref{tensoranomaly}) represent our
final result for the gauge anomalies. From this calculation one sees that
at least two powers of the YM field-strength tensor $G$ should be present in the
form $\Phi_{2n}$ to generate anomalies with respect to the $\xi$ gauge transformations.

We shall also consider a series of invariant forms that are linear in the YM field-strength tensor
$\Omega_{2n}$ given in (\ref{linearcombination}).
As we shall see, none of them generate anomalies of the standard gauge transformations, but create
anomalies with respect to the tensor gauge transformations.
We start with the invariant form that can be constructed in six dimensions
\be
\Omega_6= Tr(G  G_4 ) = d \chi_5;
\ee
the calculation proceeds as follows
\be
\delta \Omega_6 = d~ Tr(G_4 \delta A + G \delta A_3),~~~
\chi_5= \int_{0}^{1} dt Tr(G_{4t} A + G_t A_3)
\ee
thus
\be
\chi_5 = {1\over 2} Tr(G_4 A + G A_3 -A^2 A_3)=Tr(G A_3)
\ee
and its gauge variation is
$
\delta \chi_5 = d ~\chi^1_4
$
with
\be
\chi^1_4 = Tr(  G \zeta_2).
\ee
The next invariant is in eight dimensions
\be
\Omega_8= Tr(G  G_6 + G_4 G_4)  = d \chi_7 ;
\ee
the calculation proceeds again as before:
\be
\delta \Omega_8 = d~ Tr(G_6 \delta A + 2 G_4 \delta A_3 +  G \delta A_5),~~~
\chi_7= \int_{0}^{1} dt Tr(G_{6t} A + 2G_{4t} A_3 +  G_t A_5) ,
\ee
where
\beqa
\chi_7 =  {1\over 2}Tr(G_6 A+ 2 G_4 A_3 + G A_5  -2 A A^2_3-A^2 A_5)
=Tr(G A_5 +  G_4 A_3  ),
\eeqa
and its gauge variation is
$
\delta \chi_7 = d~Tr(G \zeta_4 + G_4 \zeta_2)=d\chi^1_6
$
with
\be
\chi^1_6 = Tr(G \zeta_4 + G_4 \zeta_2).
\ee
The next invariant form is in ten dimensions
\be
\Omega_{10}= Tr(G  G_8 + 3 G_4 G_6)=d \chi_9 ,
\ee
and as before:
\beqa
\delta \Omega_{10} = d~ Tr(G_8 \delta A + 3 G_6 \delta A_3 + 3 G_4 \delta A_5 +  G \delta A_7),~~~\nn\\
\chi_9= \int_{0}^{1} dt Tr(G_{8t} A + 3G_{6t} A_3 + 3G_{4t} A_5+  G_t A_7),
\eeqa
where
\beqa
\chi_9 = &{1 \over 2} Tr(G_8 A +3G_6 A_3 +3 G_4 A_5 + G A_7- A^2 A_7 -3 A \{A_3, A_5\}
-2 A^3_3)=\nn\\ &=Tr(G A_7 +2 G_4 A_5 + G_6 A_3)
\eeqa
and its gauge variation is
$
\delta \chi_9 = d~Tr(G \zeta_6 +2 G_4 \zeta_4 + G_6 \zeta_2)=d\chi^1_8
$
with
\be
 \chi^1_8 =  Tr(G \zeta_6 +2 G_4 \zeta_4 + G_6 \zeta_2).
 \ee
Finally, in twelve dimensions we have
\be
\Omega_{12} =Tr(G  G_{10} + 4 G_4 G_8 +3 G_6 G_6) =d \chi_{11},
\ee
so that
\be
\chi_{11} =Tr(G A_9 +3 G_4 A_7 + A_3 G_8 +3 G_6 A_5 )
\ee
and its  gauge variation is
$
\delta \chi_{11} = d~Tr(G \zeta_8 + 3 G_4 \zeta_6 + 3 G_6 \zeta_4 + G_8 \zeta_2),
$
thus the rank-2,4,6,10 anomalies in ten dimensions are
\be
 \chi^1_{10} =  Tr(G \zeta_8 +  3 G_4 \zeta_6 + 3 G_6 \zeta_4 + G_8 \zeta_2).
\ee
The general form of these invariants can be written as
\be
\Omega_{2n+2} = Tr(G G_{2n} + \alpha_1 G_4 G_{2n-2} +\alpha_2 G_6 G_{2n-4}+....)= d \chi_{2n+1}
\ee
where
\be
\chi_{2n+1}= Tr(G A_{2n-1}+\beta_1 G_4 A_{2n-3}+\beta_2 G_{6} A_{2n-5}+...)~.
\ee
and $\alpha_i$, $\beta_i$ are certain numerical coefficients.
The forms $\chi_{2n+1}$ are defined in  $\CD=2n+1=5,7,9,11,\dots$ dimensions.
Their gauge variation is of the form
$
\delta \chi_{2n+1}= d \chi^1_{2n},
$
where the anomalies are only with respect to the tensor gauge transformations
\be
\chi^1_{2n} =Tr(G \zeta_{2n-2}+\gamma_1 G_4 \zeta_{2n-4}+\gamma_2 G_{6} \zeta_{2n-6}+...)~,
\ee
{\em i.e.} there are no terms depending on $\xi$.

\section{\it Anomalies in 2,4,6,8, and 10 Dimensions }

Having in hands the results of the previous sections we can consider
the classification of all possible forms and anomalies in lower-dimensions.

\underline{In four dimensions} the only possible density is the standard 4-form $\CP_4=Tr(G^2)=d \omega_3$ which generates the
two-dimensional anomaly $\omega^1_{2}$  by gauge variation
$\delta \omega_3 = d \omega^1_{2}$, so that
\be
\omega^1_{2} = Tr(\xi ~ d A  ).
\ee

\underline{In six dimensions} we get two densities, $\CP_6=Tr(G^3)=d \omega_5$
and $\Phi_6 = Tr(G G_4)  =d \psi_{5}  $ which are generating
the standard non-Abelian anomaly in four dimensions $\omega^1_{4}$, as well as a new
anomaly $ \psi^1_{4}$
\beqa\label{anomalyin4}
\omega^1_{4} = Tr(\xi ~d (A d A + {1\over 2} A^3)),~~~~~~\psi^1_{4} = Tr(\zeta_2~ d A ).
\eeqa
The second anomaly $\psi^1_{4}$ is associated with the breaking of the symmetry with respect to the
gauge transformations generated by the antisymmetric tensor gauge parameter
$\zeta_2 =L^a \zeta^a_{\sigma_1 \sigma_2}dx^{\sigma_1} \wedge dx^{\sigma_2}$. In this article
we are mostly interested in classifying anomalies associated with the breaking of the Yang-Mills
gauge symmetry generated by the scalar gauge parameter $\xi = L^a \xi^a$. Therefore, from
(\ref{anomalyin4}) we
conclude that there is no new anomaly in four dimensions associated with
$\xi$ transformations except the standard one $\omega^1_{4}(\xi,A)$.

\underline{In eight dimensions   } we get three densities, $\CP_8=Tr(G^4)=d \omega_7$,
$\Phi_8 = Tr(G^2 G_4)  =d \psi_{7}  $ and $\Xi_8= Tr(G  G_6 + G_4 G_4)=d \phi_{7}$.
Therefore in six dimensions we get the standard anomaly $\omega^1_{6}$ and two new
ones $ \psi^1_{6}$ and $\phi^1_{6}$
\beqa\label{anomalyin6}
\omega^1_{6}&=&Tr(\xi ~d (A d A d A + {6\over 5}A^3 d A + {1\over 5}A^5))\nn\\
\psi^1_{6} &=& Tr(\xi~ d (d A_3 A  +d A A_3  + {1\over 2}( A^2 A_3 +A A_3 A + A_3 A^2))
+ Tr(\zeta_2~d (A d A + {1\over 2} A^3 )),~~\nn\\
   \phi^1_{6} &=&Tr( \zeta_2~G_4 + \zeta_4~G ).
\eeqa
The form $\phi^1_{6}$ describes anomalies with respect to the rank-2 $\zeta_2$ and rank-4
$\zeta_4$ gauge transformations. The form $\psi^1_{6}$ has two contributions, the last term is
associated with the $\zeta_2$ gauge transformations and the first term is clearly
associated with standard gauge transformations.
If one represents the standard gauge anomaly $\omega^1_{6}$ in  momentum representation
symbolically as
$$
\omega^1_{6} ~~~\propto ~~~Tr(L^a)^4  ~
\epsilon^{\mu_1 ... \mu_6}  e^{(1)}_{\mu_1}e^{(2)}_{\mu_2}e^{(3)}_{\mu_3} k^{(1)}_{\mu_4}k^{(2)}_{\mu_5}k^{(3)}_{\mu_6}+...,
$$
then the new  anomaly takes the form
\be
\psi^1_{6} ~~~\propto  ~~~Tr(L^a)^3 ~\epsilon^{\mu_1 ... \mu_6} e^{(1)}_{\mu_1}e^{(2)}_{\mu_2 \mu_3 \mu_4}
k^{(1)}_{\mu_5}k^{(2)}_{\mu_6}+...
\ee
where $k_{\mu_{i}}$   and  $e_{\mu_{i}}, e_{\mu_1\mu_2\mu_3}$ denote the momenta and the polarization vectors and
tensors of the corresponding gauge bosons. From this calculation one sees again that
at least two powers of the YM field-strength tensor $G$ should be present in the
form  to generate anomalies with respect to the $\xi$ gauge transformations. The order of traced generators
drops from four to three $Tr(L^a)^4 \rightarrow Tr(L^a)^3 $ and of the momenta polynomials also  drops from
three to two $(k)^3 \rightarrow (k)^2$. In higher dimensions we shall encounter the same pattern.

\underline{In ten dimensions} we get four densities, $\CP_{10}=Tr(G^5)=d \omega_9$,
$\Phi_{10} = Tr(G^3 G_4)  =d \psi_{9}  $, $\Xi_{10} = Tr(G^2 G_6 +2 G G^{2}_{4})  =d \phi_{9}  $
and $\Upsilon_{10}= Tr(G  G_8 + 3 G_4 G_6)=d \varrho_{9}$. Thus in eight dimensions we get
 $\omega^1_{8}$ and the additional three gauge anomalies
 $ \psi^1_{8}$, $ \phi^1_{8}$ and $\varrho^1_{8}$.

\underline{In twelve dimensions } we get five densities, $\CP_{12}=Tr(G^6)=d \omega_{11}$,
$\Phi_{12} = Tr(G^4 G_4)  =d \psi_{11}  $, $\Xi_{12} = Tr(G^3 G_6 +2 G^2 G^2_{4}  + G G_4 G G_4)  =d \phi_{11}  $,
$\Upsilon_{12} = Tr(G^2 G_8 +3 G  G_{4} G_6 + 3 G  G_6 G_{4} +2  G^3_4  )  =d \varrho_{11}  $
and $\Omega_{12}=Tr(G   G_{10} + 4 G_4 G_8 +3 G^2_6)=d \chi_{11}$. Thus, in ten dimensions we get
besides $\omega^1_{10}$, four additional gauge anomalies
 $ \psi^1_{10}$, $ \phi^1_{10}$, $ \varrho^1_{10}$ and $\chi^1_{10}$.
The general structure, properties  and comparison of these anomalies with
the standard one will be presented in a separate publication.

\section*{\it Acknowledgements}
One of us G.S. would like to thank Ludwig Faddeev and Luis Alvarez-Gaume
for stimulating discussions and   CERN Theory Division, where part of this work was completed,
for hospitality. This work was supported in part by the European Commission
under the ERC Advanced Grant 226371, the contract PITN-GA-2009-237920 and
by the General Secretariat for Research and Technology of Greece and
from the European Regional Development Fund (NSRF 2007-13 ACTION, KRIPIS).

\vspace{2cm}
\section{ \it Appendix A.   Gauge Transformations and Bianchi Identities  }

The gauge transformations of non-Abelian tensor gauge fields
were defined in \cite{Savvidy:2005fi,Savvidy:2005zm,Savvidy:2005ki}:
\beqa\label{fieldstrengthtensortransfor}
\delta A&=&  D \xi  ,  \\
\delta A_3&=& D \zeta_2 + [A_3, \xi]  \nonumber\\
\delta A_5&=& D \zeta_4 + 2[A_3 ,\zeta_2]+  [A_5, \xi] ,\nn\\
\delta A_7&=& D \zeta_6 + 3[A_3 ,\zeta_4]+ 3 [A_5, \zeta_2] + [A_7,\xi] ,\nn\\
\delta A_9&=& D \zeta_8 + 4[A_3 ,\zeta_6]+ 6 [A_5, \zeta_4] +4[A_7, \zeta_2] + [A_9,\xi] ,\nn\\
...........&&................................... \nn
\eeqa
where $DA_{2n+1}=dA_{2n+1}+\{A,A_{2n+1}\}$ and the corresponding field-strength tensors are
\beqa
G  &=& d A  + A^{2},  \\
 G_4 &=& d A_3 +\{ A  , A_3 \},\nn\\
G_6 &=&d A_5 + \{A, A_5\} + \{A_3, A_3\},\nn\\
G_8 &=& d A_7 + \{A, A_7\} +3 \{A_3, A_5\},\nn\\
G_{10} &=& dA_9 + \{ A, A_9 \} +4 \{ A_3, A_7\}+ 3 \{A_5, A_5\},\nn\\
...........&&................................... \nn
\eeqa
The gauge transformation (\ref{fieldstrengthtensortransfor})
of field-strength tensors is homogeneous
\beqa
\delta G &=& [G,\xi],\\
\delta G_4 &=& [G_4,\xi] +[G,\zeta_2],\nn\\
\delta G_6 &=& [G_6,\xi] + 2[G_4,\zeta_2] +[G,\zeta_4],\nn\\
\delta G_8 &=& [G_8,\xi] + 3[G_6,\zeta_2] +3[G_4,\zeta_4]+[G,\zeta_6],\nn\\
\delta G_{10} &=& [G_{10},\xi] + 4[G_8,\zeta_2] +6[G_6,\zeta_4]+4 [G_4,\zeta_6] +[G,\zeta_8],\nn\\
...........&&................................... \nn
\eeqa
The Bianchi identities  are  given by
\beqa
D G  &=& 0,\\
DG_4 + [A_3, G]&=&0,\\
D G_6 +2[A_3, G_4] + [A_5,G] &=&0,\\
D G_8 +3 [A_3, G_6] + 3 [A_5,G_4]+ [A_7,G] &=&0,\\
D G_{10} +4 [A_3, G_8] + 6 [A_5,G_6]+ 4 [A_7,G_4]+[A_9,G]  &=&0,\nn\\
.........................................&&..... \nn
\eeqa
where $DG_{2n}  = dG_{2n}  + [A , G_{2n} ]$. Generalizing  Zumino's construction \cite{Zumino:1983ew},
we introduce a one-parameter family of potentials and field-strengths as :
\beqa
A_{t}= t A,~~~A_{3t}= t A_3,~~~A_{5t}= t A_5,~~~A_{7t}= t A_7,~~~A_{9t}= t A_9,\nn\\
G_{t}= t G +(t^2-t)A^{2},\nn\\
G_{4t}= t G_4 +(t^2-t)\{A,A_{3}\},\nn\\
G_{6t}= t G_6 +(t^2-t)(\{A,A_{5}\}+ \{A_3,A_{3}\}) ,\nn\\
G_{8t}= t G_8 +(t^2-t)(\{A,A_{7}\}+ 3\{A_3,A_{5}\}) ,\nn\\
G_{10t}= t G_{10} +(t^2-t)(\{A,A_{9}\}+ 4\{A_3,A_{7}\}+ 3\{A_5,A_{5}\} ) ,\nn\\
............................................. \nn
\eeqa
In order to find out the secondary forms it is useful to perform the variation
of the fields in the corresponding expressions
\beqa
\delta G &=& D (\delta A) ,\nn\\
\delta G_4 &=& D(\delta A_3) + \{A_3,\delta A \},\nn\\
\delta G_6 &=& D(\delta A_5) + \{A_5,\delta A \}+ 2\{A_3,\delta A_3 \},\nn\\
\delta G_8 &=& D(\delta A_7) + \{A_7,\delta A \}+ 3\{A_5,\delta A_3 \}+ 3\{A_3,\delta A_5 \},\nn\\
\delta G_{10} &=& D(\delta A_9) + \{A_9,\delta A \}+ 4\{A_7,\delta A_3 \}+ 6\{A_5,\delta A_5 \}+ 4\{A_3,\delta A_7 \},\nn\\
..........&=&................................... \nn
\eeqa

\appendix

\vfill
\end{document}